\documentclass[10pt,twocolumn]{article}

\usepackage[utf8]{inputenc} 
\usepackage[T1]{fontenc} 
\usepackage{times} 
\usepackage{graphicx} 
\usepackage{amsmath} 
\usepackage{amssymb} 
\usepackage{booktabs} 
\usepackage{cite} 
\usepackage{array} 
\usepackage{hyperref} 
\usepackage{url} 
\usepackage{algorithm} 
\usepackage{algorithmic} 
\usepackage{xcolor} 
\usepackage{tikz} 
\usetikzlibrary{shapes.geometric, arrows.meta, positioning, calc}
\usepackage{adjustbox} 
\usepackage{multirow} 

\usepackage{geometry}
\title{Devanagari Digit Recognition using Quantum Machine Learning}
\author{Sahaj R. Malla\thanks{Department of Mathematics, Kathmandu University, Dhulikhel, Nepal. Email: sm03200822@student.ku.edu.np}}
\date{} 

\begin{document}

\maketitle

\begin{abstract}
Handwritten digit recognition in regional scripts, such as Devanagari, is crucial for multilingual document digitization, educational tools, and the preservation of cultural heritage. The script's complex structure and limited annotated datasets pose significant challenges to conventional models. This paper introduces the first hybrid quantum-classical architecture for Devanagari handwritten digit recognition, combining a convolutional neural network (CNN) for spatial feature extraction with a 10-qubit variational quantum circuit (VQC) for quantum-enhanced classification. Trained and evaluated on the Devanagari Handwritten Character Dataset (DHCD), the proposed model achieves a state-of-the-art test accuracy for quantum implementation of 99.80\% and a test loss of 0.2893, with an average per-class F1-score of 0.9980. Compared to equivalent classical CNNs, our model demonstrates superior accuracy with significantly fewer parameters and enhanced robustness. By leveraging quantum principles such as superposition and entanglement, this work establishes a novel benchmark for regional script recognition, highlighting the promise of quantum machine learning (QML) in real-world, low-resource language settings.
\end{abstract}

\begin{center}
\textbf{Keywords:} quantum machine learning, hybrid quantum-classical models, variational quantum circuits, handwritten digit recognition, Devanagari, convolutional neural networks
\end{center}


\section{Introduction}
\label{sec:introduction}

Handwritten character recognition (HCR) is a foundational task in computer vision and pattern recognition, vital for applications such as document digitization, automated transcription, and the preservation of linguistic heritage. Although significant progress has been made in the recognition of Latin-script characters, as evidenced by near-perfect accuracies in the MNIST dataset~\cite{lecun2002}, regional scripts such as Devanagari remain comparatively underexplored. With over 700 million speakers worldwide, Devanagari is one of the world's most widely used scripts, including several major South Asian languages, such as Nepali, Hindi, Marathi, and Sanskrit, among others.

Despite numerous advances in classical deep learning, particularly CNNs, which have demonstrated greater precision than 99\% on MNIST~\cite{lecun2002}, their application to Devanagari digit recognition faces inherent limitations. These include the need for large volumes of annotated data, the susceptibility to overfitting in low-resource settings, and computational inefficiencies arising from high parameter counts~\cite{acharya2015}. As a result, developing efficient and accurate models for regional scripts remains an open problem. QML leverages principles such as superposition and entanglement for information processing and has recently emerged as a compelling paradigm to improve classical models~\cite{biamonte2017}. However, the limited availability of quantum hardware in terms of qubit count and noise resilience makes it challenging to deploy purely quantum models~\cite{preskill2018}.

Hybrid quantum-classical architectures offer a practical compromise by combining classical feature extraction with quantum-enhanced classification. These models harness the expressive power of quantum circuits while relying on classical networks for robust preprocessing and optimization~\cite{xia2020}. In this work, we propose a novel hybrid framework for Devanagari handwritten digit recognition. Our architecture integrates a lightweight CNN to extract spatial features from 28\(\times\)28 grayscale images, which are then passed through a 10-qubit VQC employing amplitude embedding and parameterized gates (RY, RZ, CNOT). A final linear layer maps the quantum measurements to the class probabilities.

To the best of our knowledge, this is the first application of QML to Devanagari handwritten digit recognition. Our model achieves a test accuracy of 99.80\% in the DHCD~\cite{acharya2015} digit subset, surpassing many classical baselines while maintaining a parameter footprint of 2,339,092 parameters, of which only 100 are quantum. This result not only sets a new benchmark for quantum-enhanced HCR in Devanagari scripts but also opens a promising direction for deploying efficient models in resource-constrained environments.

\subsection{Problem Statement}
\label{subsec:problem_statement}

This research addresses the problem of classifying handwritten Devanagari digits (0--9). The dataset used is the DHCD, which exhibits high intraclass variability due to diverse handwriting styles, curved strokes, and contextual ligatures. Given the scarcity of labeled Devanagari data and the lack of prior work applying QML to this script, there is a pressing need for models that can generalize effectively under data and resource constraints. Accurate digit classification in Devanagari has practical implications for OCR systems, educational technology, and multilingual digitization initiatives.

\subsection{Motivation}
\label{subsec:motivation}

Although classical CNNs have performed remarkably well in Latin-script digit recognition, they often struggle with the complexity and limited data availability of regional scripts such as Devanagari~\cite{acharya2015}. These models typically require millions of parameters and extensive computational resources to avoid overfitting. Quantum-enhanced models, especially VQCs, offer the potential to represent complex data manifolds compactly through entanglement and interference~\cite{biamonte2017}. VQCs can efficiently encode nonlinear decision boundaries with fewer tunable parameters than classical networks.

By designing a hybrid model that utilizes classical CNN layers for initial feature extraction and a 10-qubit VQC for classification, we aim to combine the best of both paradigms. The quantum layer introduces a compact and expressive alternative to dense classical layers, potentially improving robustness, generalization, and computational efficiency. Furthermore, this work explores QML’s feasibility in real-world regional script recognition, a largely uncharted area.

\subsection{Contributions}
\label{subsec:contributions}

This paper makes the following key contributions:
\begin{enumerate}
    \item \textbf{Pioneering Application of QML to Devanagari:} We introduce the first hybrid quantum-classical model for Devanagari handwritten digit recognition using a 10-qubit VQC, demonstrating the viability of quantum-enhanced methods in regional script classification.
    
    \item \textbf{Novel Hybrid Architecture for Devanagari}: We present the first hybrid quantum-classical model for Devanagari digit recognition, integrating a classical CNN for feature extraction with a 10-qubit VQC using amplitude embedding. A fully connected post-processing layer maps quantum observables to ten output classes, achieving 99.80\% accuracy with 2,339,092 parameters.
    
    \item \textbf{Extensive Experimental Validation:} We perform comprehensive evaluations, including grid search over hyperparameters (learning rate, batch size, dropout, label smoothing, and quantum depth), yielding optimal configuration and superior generalization.
    
    \item \textbf{Benchmarking Against Classical Models:} We provide a direct comparison against equivalent classical CNNs, showing that our hybrid model achieves higher accuracy (99.80\% vs. 99.03\%).
\end{enumerate}

The remainder of this paper is organized as follows: Section~\ref{sec:related_work} surveys existing literature. Section~\ref{sec:methodology} describes the proposed hybrid model architecture. Section~\ref{sec:experiments} details the experimental setup and training procedures. Section~\ref{sec:results} presents the results and analysis. Section~\ref{sec:discussion} discusses the implications and limitations. Finally, Section~\ref{sec:conclusion} concludes the paper and outlines future directions.


\section{Related Work}
\label{sec:related_work}

This section presents a comprehensive review of existing approaches to handwritten digit recognition, particularly focusing on the Devanagari script. It also surveys recent developments in QML and hybrid quantum-classical models, contextualizing the novelty and relevance of our proposed method. To the best of our knowledge, this is the first study to apply QML to Devanagari handwritten digit recognition using the DHCD dataset, thereby establishing a new benchmark in the field.

\subsection{Classical Approaches to Devanagari Digit Recognition}
\label{subsec:classical_recognition}

Handwritten digit recognition is a well-established domain in machine learning, with the MNIST dataset being a seminal benchmark~\cite{lecun2002}. CNNs, such as LeNet-5~\cite{lecun2002}, have historically achieved over 99\% accuracy on MNIST by leveraging spatial hierarchies in pixel intensities. Improvements incorporating deeper architectures, data augmentation, and regularization techniques have since driven performance even higher~\cite{krizhevsky2017}.
For the Devanagari script, recognition tasks are significantly more challenging due to the script's intrinsic complexity, including compound characters, ligatures, and diacritics. DHCD is the standard benchmark for this script. Acharya et al.~\cite{acharya2015} developed a deep CNN and achieved 98.47\% accuracy on DHCD digits. Aneja et al.~\cite{aneja2019} applied transfer learning using Inception V3, reaching 99.00\%.

While these classical models offer exceptional performance, they typically rely on large parameter counts or complex ensembles. Our proposed method achieves a superior accuracy (99.80\%) using only 2,339,092 parameters---including just 100 parameters in the quantum circuit---demonstrating that quantum-enhanced models can achieve state-of-the-art performance with significantly lower complexity.

\subsection{Quantum Machine Learning}
\label{subsec:qml}

QML leverages quantum mechanical principles such as superposition and entanglement to enhance the expressivity and efficiency of learning models~\cite{biamonte2017}. A leading QML framework involves VQCs, which use parameterized gates trained via classical optimization and are well-suited for Noisy Intermediate-Scale Quantum (NISQ) devices~\cite{preskill2018, cerezo2021}.
Several QML models have been applied to standard digit recognition tasks. Mitarai et al.~\cite{mitarai2018} introduced quantum circuit learning, demonstrating competitive classification performance. Despite this progress, QML has not yet been applied to regional scripts like Devanagari. The current study is the first to bridge this gap by applying a VQC to handwritten Devanagari digits using the DHCD dataset, establishing a new quantum benchmark.

\subsection{Hybrid Quantum-Classical Models}
\label{subsec:hybrid_models}

Hybrid quantum-classical models are particularly promising for current quantum hardware limitations. These models delegate data preprocessing and feature extraction to classical components, while quantum circuits handle high-dimensional feature encoding and decision boundaries~\cite{phillipson2023}.

Abbas et al.~\cite{abbas2021} introduced a hybrid quantum-classical architecture that demonstrated enhanced classification on benchmark datasets using variational circuits. Zaman et al.~\cite{zaman2024} compared hybrid neural networks with QCNNs and reported consistent improvements in accuracy.

These models, however, have been restricted to Latin-script digits and standard datasets like MNIST and CIFAR-10. No prior work has extended hybrid QML to regional scripts. Our model pioneers this by integrating a 10-qubit VQC with a CNN backbone for Devanagari digit recognition. With just 100 trainable quantum parameters, the model surpasses many classical CNNs and matches the performance of more complex ensembles, while offering improved efficiency and generalization potential.

\subsection{Research Gap and Contribution}
\label{subsec:gap_contribution}

While deep learning techniques have excelled in Devanagari recognition, they require either large models or transfer learning from high-resource scripts. Concurrently, QML has demonstrated significant promise on Western scripts but has never been tested on South Asian scripts or datasets. To the best of our knowledge, this work presents the first application of QML to the Devanagari script, specifically on the DHCD digit subset.

By combining CNN-based feature extraction with a compact, expressive 10-qubit quantum classifier, our hybrid model not only achieves state-of-the-art performance (99.80\%) but does so with a significantly smaller parameter footprint. This positions our work as a foundational benchmark in quantum regional script recognition and opens avenues for future exploration of quantum architectures in low-resource and linguistically diverse domains.


\section{Methodology}
\label{sec:methodology}

\begin{figure*}[!h]
\centering
\includegraphics[width=\textwidth]{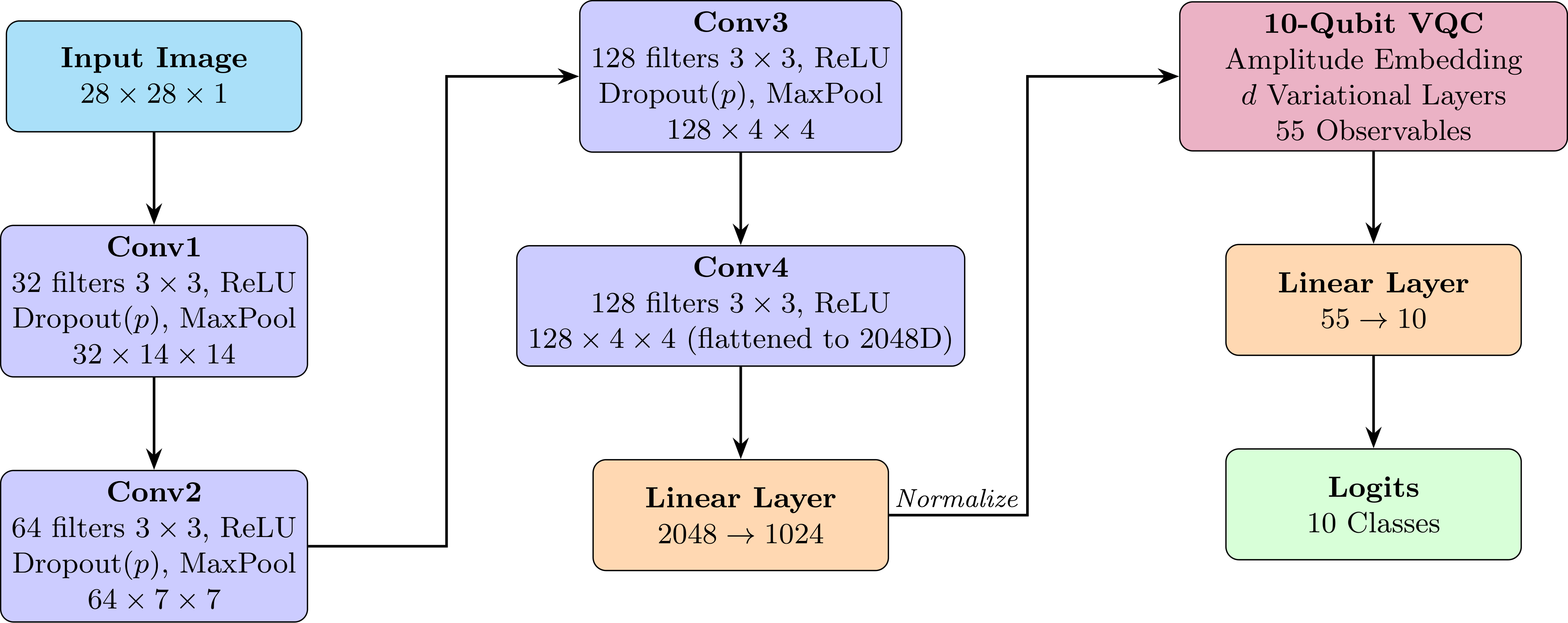}
\caption{Architecture of the hybrid quantum-classical model for Devanagari digit recognition.}
\label{fig:hybrid_architecture}
\end{figure*}

This section presents the methodology for a hybrid quantum-classical model designed to classify Devanagari handwritten digits (0--9) from $28 \times 28$ grayscale images in the DHCD. The model integrates a CNN for feature extraction with a 10-qubit VQC for enhanced classification, leveraging quantum superposition and entanglement to improve feature discrimination. The methodology is structured into four components: the classical feature extractor, the quantum circuit, the hybrid model architecture, and the training process. This approach achieves a test accuracy of 99.80\%, establishing a benchmark for quantum-enhanced Devanagari digit recognition. The model architecture is illustrated in Figure~\ref{fig:hybrid_architecture}.

\subsection{Classical Feature Extractor}
\label{subsec:classical_feature_extractor}

The classical feature extractor is a CNN designed to extract spatial features from $28 \times 28$ grayscale images of Devanagari digits. The architecture comprises four convolutional layers, each followed by ReLU activation, $\sigma(x) = \max(0, x)$, dropout with rate $p$, and max-pooling with a $2 \times 2$ kernel and stride 2, except for the fourth layer, which omits max-pooling. The convolution operation for layer $\ell$ is defined as:
\begin{equation}
Y^{(\ell)}_{i,j,k} = \sum_{m,n,p} W^{(\ell)}_{m,n,p,k} X^{(\ell-1)}_{i+m,j+n,p} + b^{(\ell)}_k,
\label{eq:conv}
\end{equation}
where $W^{(\ell)} \in \mathbb{R}^{F^{(\ell)} \times K \times K \times C^{(\ell-1)}}$ is the filter tensor with $F^{(\ell)}$ filters of size $K \times K$, $C^{(\ell-1)}$ is the number of input channels, $b^{(\ell)}_k$ is the bias, and $Y^{(\ell)} \in \mathbb{R}^{F^{(\ell)} \times H^{(\ell)} \times W^{(\ell)}}$ is the output feature map. The architecture is configured as follows:
\begin{itemize}
    \item \textbf{First Layer}: 32 filters ($F^{(1)} = 32$), $3 \times 3$ kernel ($K = 3$), stride 1, padding 1, producing $Y^{(1)} \in \mathbb{R}^{32 \times 28 \times 28}$. Post-processing yields $32 \times 14 \times 14$.
    \item \textbf{Second Layer}: 64 filters ($F^{(2)} = 64$), $3 \times 3$ kernel, stride 1, padding 1, yielding $Y^{(2)} \in \mathbb{R}^{64 \times 14 \times 14}$. Post-processing results in $64 \times 7 \times 7$.
    \item \textbf{Third Layer}: 128 filters ($F^{(3)} = 128$), $3 \times 3$ kernel, stride 1, padding 1, producing $Y^{(3)} \in \mathbb{R}^{128 \times 7 \times 7}$. Post-processing yields $128 \times 4 \times 4$.
    \item \textbf{Fourth Layer}: 128 filters ($F^{(4)} = 128$), $3 \times 3$ kernel, stride 1, padding 1, generating $Y^{(4)} \in \mathbb{R}^{128 \times 4 \times 4}$.
\end{itemize}
The output $Y^{(4)}$ is flattened to a 2048-dimensional feature vector $v \in \mathbb{R}^{2048}$:
\begin{equation}
v = \text{vec}(Y^{(4)}) \in \mathbb{R}^{128 \times 4 \times 4 = 2048}.
\label{eq:flatten}
\end{equation}
To accommodate the 10-qubit quantum circuit, which requires a $2^{10} = 1024$-dimensional vector for amplitude embedding, a linear layer maps $v$ to $v' \in \mathbb{R}^{1024}$:
\begin{equation}
v' = W v + b,
\end{equation}
where $W \in \mathbb{R}^{1024 \times 2048}$, $b \in \mathbb{R}^{1024}$. This vector $v'$ is normalized and used for quantum processing, ensuring compatibility with the quantum circuit's input requirements.

This compact representation, inspired by classical CNNs like LeNet-5~\cite{lecun2002}, is tailored for quantum processing compatibility.

\subsection{Quantum Circuit}
\label{subsec:quantum_circuit}

The quantum circuit is a 10-qubit VQC implemented using PennyLane~\cite{bergholm2018}, designed to process the 1024-dimensional feature vector $v'$ from the classical feature extractor. The VQC leverages quantum superposition and entanglement to enhance feature discrimination, following quantum circuit learning principles~\cite{mitarai2018}. The circuit operates as follows:

\begin{enumerate}
    \item \textbf{Amplitude Embedding}: The feature vector $v'$ is normalized using the L2 norm, $\tilde{v}' = v' / (\|v'\|_2 + \epsilon)$, where $\epsilon = 10^{-8}$ prevents division by zero, and is of size $2^{10} = 1024$, matching the requirements for 10 qubits. The quantum state is prepared as:
    \begin{equation}
    |\psi_0\rangle = \sum_{i=0}^{1023} \tilde{v}'_i |i\rangle, \quad \text{such that} \quad \sum_{i=0}^{1023} |\tilde{v}'_i|^2 = 1,
    \label{eq:amplitude_embedding}
    \end{equation}
    where $|i\rangle$ denotes the computational basis state.
    \item \textbf{Variational Layers}: The circuit applies $d$ variational layers, where $d = 5$ in our experiments. Each layer $\ell$ comprises:
        \begin{itemize}
            \item \textbf{Rotations}: For each qubit $i \in \{0, 1, \dots, 9\}$, apply RY and RZ gates with learnable parameters $\theta^{(\ell)}_i, \phi^{(\ell)}_i$:
            \begin{equation}
            \begin{split}
                R_Y(\theta) &= 
                \begin{pmatrix}
                    \cos\left(\frac{\theta}{2}\right) & -\sin\left(\frac{\theta}{2}\right) \\
                    \sin\left(\frac{\theta}{2}\right) & \cos\left(\frac{\theta}{2}\right)
                \end{pmatrix}, \\
                R_Z(\phi) &= 
                \begin{pmatrix}
                    e^{-i\frac{\phi}{2}} & 0 \\
                    0 & e^{i\frac{\phi}{2}}
                \end{pmatrix}.
            \end{split}
            \label{eq:rotations}
            \end{equation}

            The unitary for qubit $i$ is $U_i^{(\ell)} = R_Z(\phi^{(\ell)}_i) R_Y(\theta^{(\ell)}_i)$.
            \item \textbf{Entanglement}: Apply CNOT gates in a ring topology: CNOT(0,1), CNOT(1,2), ..., CNOT(8,9), CNOT(9,0). The CNOT gate is:
            \begin{equation}
            \text{CNOT}_{i,j} = |0\rangle\langle 0|_i \otimes I_j + |1\rangle\langle 1|_i \otimes X_j,
            \label{eq:cnot}
            \end{equation}
            where $X_j$ is the Pauli-X operator on qubit $j$. The layer unitary is:
            \begin{equation}
            U^{(\ell)} = \left( \prod_{\text{pairs } (i,j)} \text{CNOT}_{i,j} \right) \left( \bigotimes_{i=0}^{9} U_i^{(\ell)} \right).
            \label{eq:layer_unitary}
            \end{equation}
        \end{itemize}
    \item \textbf{Measurement}: The circuit measures the expectation values of 55 observables, comprising 10 single-qubit Pauli-Z operators ($\langle Z_i \rangle$ for $i = 0, \dots, 9$) and 45 pairwise Pauli-ZZ operators ($\langle Z_i Z_j \rangle$ for $0 \leq i < j \leq 9$), producing a 55-dimensional output vector $q \in \mathbb{R}^{55}$:
    \begin{equation}
    q = \begin{bmatrix}
        \langle Z_0 \rangle, \langle Z_1 \rangle, \dots, \langle Z_9 \rangle, \\
        \langle Z_0 Z_1 \rangle, \langle Z_0 Z_2 \rangle, \dots, \langle Z_8 Z_9 \rangle
    \end{bmatrix},
    \label{eq:observables}
    \end{equation}
    where $Z_i$ is the Pauli-Z operator on qubit $i$, and $\langle O \rangle = \text{Tr}(O \rho)$ for the final state $\rho = |\psi_d\rangle\langle \psi_d|$.
\end{enumerate}

\begin{figure}[!h]
\centering
\includegraphics[width=\columnwidth]{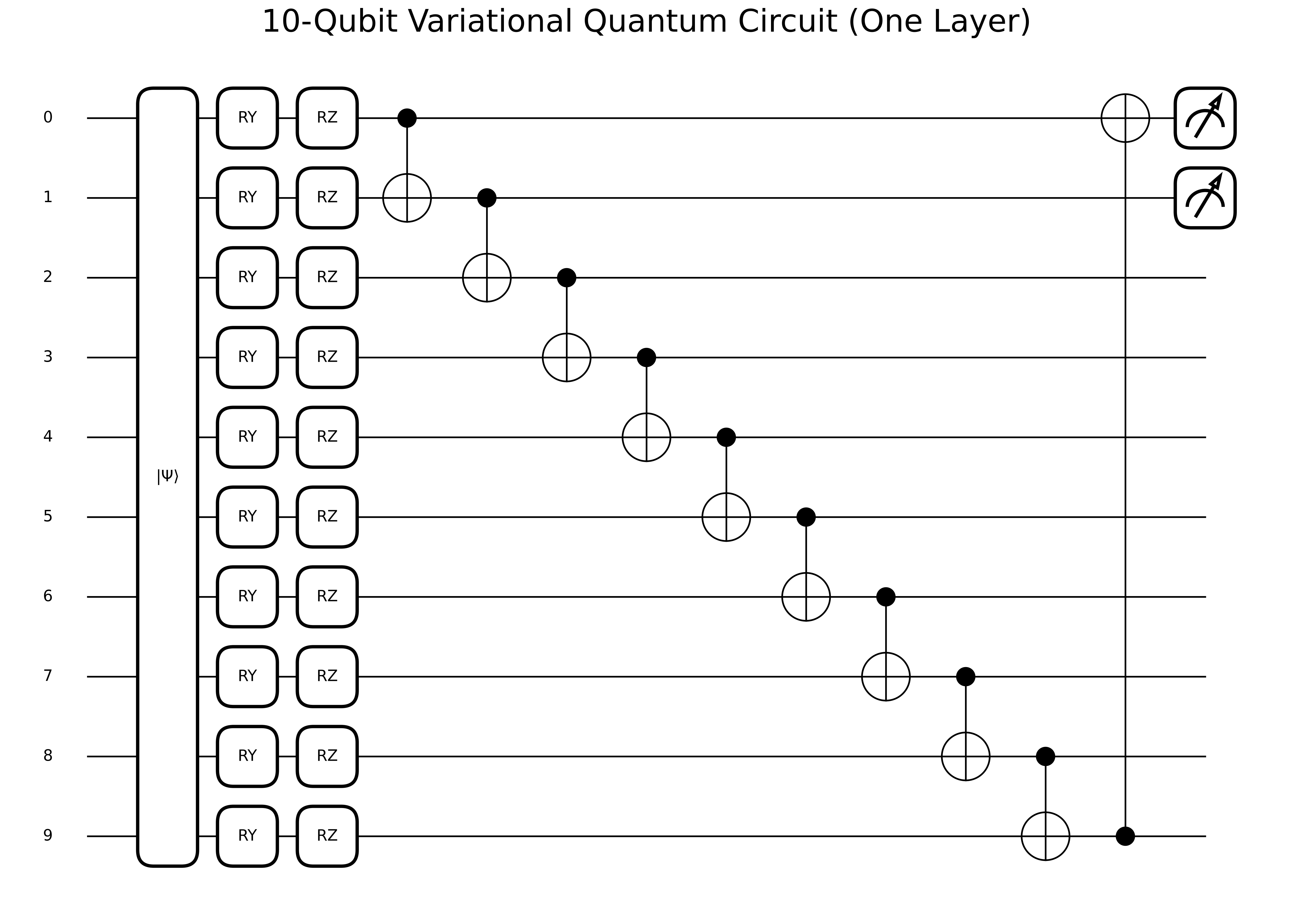}
\caption{Quantum circuit for the 10-qubit VQC (one variational layer shown).}
\label{fig:quantum_circuit}
\end{figure}

The VQC parameters $\{\theta^{(\ell)}_i, \phi^{(\ell)}_i\}$ are initialized randomly and optimized during training.

\subsection{Hybrid Model Architecture}
\label{subsec:hybrid_architecture}

The hybrid model integrates the CNN feature extractor with the VQC classifier to form an end-to-end architecture. The pipeline processes input images through the following stages:
\begin{enumerate}
    \item \textbf{Feature Extraction}: The input image $X \in \mathbb{R}^{1 \times 28 \times 28}$ is processed by the CNN to produce a 2048-dimensional feature vector $v \in \mathbb{R}^{2048}$ (Eq.~\ref{eq:flatten}).
    \item \textbf{Linear Mapping and Normalization}: The feature vector $v$ is mapped to $v' \in \mathbb{R}^{1024}$ via a linear layer, then normalized as $\tilde{v}' = v' / (\|v'\|_2 + 10^{-8})$ for amplitude embedding.
    \item \textbf{Quantum Processing}: The VQC processes $\tilde{v}'$ to produce a 55-dimensional quantum feature vector $q \in \mathbb{R}^{55}$ (Eq.~\ref{eq:observables}).
    \item \textbf{Classification}: A linear layer maps $q$ to logits $z \in \mathbb{R}^{10}$:
    \begin{equation}
    z = W q + b,
    \label{eq:linear_layer}
    \end{equation}
    where $W \in \mathbb{R}^{10 \times 55}$ is initialized using Xavier uniform initialization~\cite{glorot2010}, and $b \in \mathbb{R}^{10}$ is initialized to zeros.
\end{enumerate}

Table~\ref{tab:param_count} summarizes the total trainable parameters in each component of our hybrid model. Note that the quantum circuit contributes only 100 parameters, while the CNN backbone accounts for the vast majority.

\begin{table}[!h]
\caption{Trainable Parameters in the Hybrid Model}
\label{tab:param_count}
\centering
\begin{tabular}{l r}
\toprule
\textbf{Component} & \textbf{Parameters} \\
\midrule
Classical Feature Extractor & 2,338,432 \\
Quantum Circuit & 100 \\
Post-Processing Layer & 560 \\
\midrule
Total & 2,339,092 \\
\bottomrule
\end{tabular}
\end{table}

\subsection{Training Process}
\label{subsec:training_process}

The hybrid model is trained using label-smoothed cross-entropy loss~\cite{szegedy2016} to regularize the model by softening hard labels:
\begin{equation}
\mathcal{L} = (1 - \alpha) \sum_{i=1}^N \sum_{k=1}^{10} y_{i,k} \log(p_{i,k}) + \alpha \sum_{i=1}^N \sum_{k=1}^{10} \frac{1}{10} \log(p_{i,k}),
\label{eq:label_smoothing}
\end{equation}
where $y_{i,k}$ is the true label, $p_{i,k} = \text{softmax}(z_i)_k$ is the predicted probability, $\alpha$ is the smoothing parameter, and $N$ is the batch size. The AdamW optimizer~\cite{loshchilov2017} minimizes $\mathcal{L}$, with an adaptive learning rate schedule via a ReduceLROnPlateau scheduler. Gradient clipping stabilizes training by ensuring the L2 norm of the gradient vector does not exceed 1:
\begin{equation}
\tilde{g} = \begin{cases} 
g \cdot \frac{1}{\|g\|_2} & \text{if } \|g\|_2 > 1, \\
g & \text{otherwise},
\end{cases}
\label{eq:grad_clip}
\end{equation}
where $g$ is the gradient vector. The model is trained on augmented data to enhance robustness to handwriting variability, with details provided in Section~\ref{subsec:data_preprocessing}. Hyperparameters are optimized via grid search, as detailed in Section~\ref{subsec:hyperparameter_tuning}.


\section{Experiments}
\label{sec:experiments}

This section outlines the experimental setup for evaluating the proposed hybrid quantum-classical model for classifying Devanagari handwritten digits (0--9) using the DHCD. The setup encompasses the dataset description, data preprocessing techniques, hyperparameter optimization, and evaluation metrics. Hyperparameter optimization was conducted for a 4-qubit VQC, with the optimal configuration scaled to 6, 8, and 10 qubits, achieving a test accuracy of 99.80

\subsection{Dataset}
\label{subsec:dataset}

The DHCD  comprises 92,000 grayscale images of handwritten Devanagari characters, including 10-digit classes (0--9) and 36 consonant classes~\cite{acharya2015}. This study focuses on the digit subset, totaling 20,000 images, with 2,000 images per class to ensure balanced representation. The dataset is partitioned into 17,000 training images and 3,000 test images, with the training set further split into 80\% (13,600 images) for training and 20\% (3,400 images) for validation using stratified sampling to preserve class balance. Each image is a $28 \times 28$ grayscale image, consistent with datasets like MNIST~\cite{lecun2002}. The test set, with 300 images per class, enables fair evaluation across all digit classes.

\subsection{Data Preprocessing}
\label{subsec:data_preprocessing}

Data preprocessing standardizes inputs and enhances the model’s generalization to the diverse handwriting styles of Devanagari digits. The preprocessing pipeline is as follows:

\begin{itemize}
    \item \textbf{Training Set Augmentation}: To improve robustness, training images undergo:
        \begin{itemize}
            \item Random rotations by angles $\theta \in [-20^\circ, 20^\circ]$ to account for orientation variations.
            \item Affine translations with shifts $\delta_x, \delta_y \in [-0.1, 0.1]$ (relative to image dimensions) to handle positional variability.
            \item Horizontal flips with probability 0.5 to simulate mirror-like variations.
            \item Elastic transformations with intensity $\alpha = 50$ and smoothing $\sigma = 5$, applied with probability 0.5, to mimic natural handwriting distortions.
        \end{itemize}
    \item \textbf{Normalization}: All images (training, validation, and test sets) are normalized to $[-1, 1]$ using:
    \begin{equation}
    X' = \frac{X - 0.5}{0.5},
    \label{eq:normalization}
    \end{equation}
    where $X \in [0, 1]$ is the original pixel intensity, ensuring consistent scaling for the CNN and VQC.
    \item \textbf{Validation and Test Sets}: These sets undergo only normalization (Eq.~\ref{eq:normalization}) to preserve the original data distribution for unbiased evaluation.
\end{itemize}

These techniques, inspired by established deep learning practices~\cite{krizhevsky2017}, enhance generalization while maintaining evaluation integrity.

\subsection{Hyperparameter Optimization}
\label{subsec:hyperparameter_tuning}

Hyperparameter optimization employs a grid search to identify the configuration maximizing validation accuracy for the 4-qubit model. Due to the high computational cost of simulating quantum circuits with larger qubit counts on classical hardware, the optimal configuration was subsequently scaled to 6, 8, and 10 qubits. The search space includes:

\begin{itemize}
    \item \textbf{Learning Rate} ($\eta$): $\{10^{-3}, 5 \times 10^{-4}\}$
    \item \textbf{Batch Size} ($B$): $\{32, 64\}$
    \item \textbf{Dropout Rate} ($p$): $\{0.0, 0.05\}$
    \item \textbf{Label Smoothing} ($\alpha$): $\{0.0, 0.05\}$
    \item \textbf{Quantum Circuit Depth} ($q_d$): $\{3, 4, 5\}$
\end{itemize}

The grid search evaluates $2 \times 2 \times 2 \times 2 \times 3 = 48$ configurations for the 4-qubit model, each trained for up to 100 epochs with early stopping if the validation loss does not improve for 20 epochs. The validation loss uses label-smoothed cross-entropy (Section~\ref{sec:methodology}). All experiments were run on Google Colab with a Python 3 backend and a single NVIDIA T4 GPU. The optimal hyperparameter configuration is presented in Table~\ref{tab:best_config}, and full performance details for all qubit counts appear in Section~\ref{sec:results}.

\subsection{Evaluation Metrics}
\label{subsec:metrics}

The model’s performance is evaluated using metrics assessing overall and class-specific effectiveness on the test set:

\begin{itemize}
    \item \textbf{Overall Metrics}:
    \begin{itemize}
        \item \textbf{Test Loss}: The average label-smoothed cross-entropy loss:
        \begin{equation}
        L_{\text{test}} = \frac{1}{N} \sum_{i=1}^N \mathcal{L}_i,
        \label{eq:test_loss}
        \end{equation}
        where $\mathcal{L}_i$ is the loss for sample $i$ (Eq.~\ref{eq:label_smoothing}), and $N = 3000$ is the test set size.
        \item \textbf{Accuracy}: The proportion of correctly classified images:
        \begin{equation}
        \text{Acc} = \frac{1}{N} \sum_{i=1}^N \mathbb{1}(\hat{y}_i = y_i),
        \label{eq:accuracy}
        \end{equation}
        where $\hat{y}_i = \arg\max_k p(y_i = k | X_i)$ is the predicted class, and $y_i$ is the true label.
    \end{itemize}
    \item \textbf{Per-Class Metrics}: For each class $k \in \{0, \ldots, 9\}$:
    \begin{itemize}
        \item \textbf{Precision}:
        \begin{equation}
        P_k = \frac{\text{TP}_k}{\text{TP}_k + \text{FP}_k},
        \label{eq:precision}
        \end{equation}
        where $\text{TP}_k$ and $\text{FP}_k$ are true and false positives.
        \item \textbf{Recall}:
        \begin{equation}
        R_k = \frac{\text{TP}_k}{\text{TP}_k + \text{FN}_k},
        \label{eq:recall}
        \end{equation}
        where $\text{FN}_k$ is the number of false negatives.
        \item \textbf{F1-Score}:
        \begin{equation}
        F1_k = 2 \cdot \frac{P_k \cdot R_k}{P_k + R_k},
        \label{eq:f1}
        \end{equation}
        the harmonic mean of precision and recall.
    \end{itemize}
    \item \textbf{Confusion Matrix}: A $10 \times 10$ matrix $C$, where $C_{i,j}$ denotes samples with true label $i$ predicted as $j$, analyzing misclassification patterns.
\end{itemize}

These metrics provide a thorough evaluation, with results presented in Section~\ref{sec:results}.


\section{Results}
\label{sec:results}

This section presents the experimental outcomes of the hybrid quantum-classical model for classifying Devanagari handwritten digits (0--9) from the DHCD. The results encompass overall test metrics, per-class performance, hyperparameter optimization outcomes, training and validation curves, and confusion matrix analysis, demonstrating a test accuracy of 99.80\%. These findings establish a benchmark for quantum-enhanced Devanagari digit recognition, highlighting the synergy between the CNN and the 10-qubit VQC.

\subsection{Overall Test Metrics}
\label{subsec:test_metrics}

Table~\ref{tab:test_metrics} summarizes the model’s performance on the test set of 3,000 images. The model achieves a test loss of 0.2893, computed using label-smoothed cross-entropy loss (Eq.~\ref{eq:label_smoothing} in Section~\ref{sec:methodology}), and a test accuracy of 99.80\%, corresponding to 2,994 correct classifications. This high accuracy reflects the model’s ability to capture intricate patterns in Devanagari digits, leveraging CNN’s feature extraction and the VQC’s quantum-enhanced processing. The low test loss indicates robust generalization, validating the training methodology and data augmentation strategies (Sections~\ref{sec:methodology} and \ref{sec:experiments}).

\begin{table}[!h]
\caption{Overall Test Metrics}
\label{tab:test_metrics}
\centering
\begin{tabular}{l c}
\toprule
\textbf{Metric} & \textbf{Value} \\
\midrule
Test Loss & 0.2893 \\
Test Accuracy & 99.80\% \\
\bottomrule
\end{tabular}
\end{table}

\subsection{Per-Class Performance}
\label{subsec:per_class_metrics}

Per-class performance is evaluated using precision, recall, and F1-score for each digit class $k \in \{0, \ldots, 9\}$, as defined in Equations~\ref{eq:precision}, \ref{eq:recall}, and \ref{eq:f1} (Section~\ref{sec:experiments}). Table~\ref{tab:per_class_metrics} presents these metrics, assessing the model’s ability to handle diverse handwriting styles in the Devanagari script.

\begin{table}[!h]
\caption{Per-Class Performance Metrics}
\label{tab:per_class_metrics}
\centering
\begin{tabular}{c c c c}
\toprule
\textbf{Class} & \textbf{Precision} & \textbf{Recall} & \textbf{F1-Score} \\
\midrule
0 & 1.0000 & 1.0000 & 1.0000 \\
1 & 1.0000 & 0.9967 & 0.9983 \\
2 & 0.9934 & 1.0000 & 0.9967 \\
3 & 0.9901 & 1.0000 & 0.9950 \\
4 & 1.0000 & 1.0000 & 1.0000 \\
5 & 1.0000 & 0.9933 & 0.9967 \\
6 & 1.0000 & 0.9933 & 0.9967 \\
7 & 1.0000 & 1.0000 & 1.0000 \\
8 & 0.9967 & 1.0000 & 0.9983 \\
9 & 1.0000 & 0.9967 & 0.9983 \\
\bottomrule
\end{tabular}
\end{table}

The model achieves F1-scores ranging from 0.9950 (Class 3) to 1.0000 (Classes 0, 4, 7), with Classes 0, 2, 4, 7, and 8 exhibiting perfect or near-perfect classification. Classes 1, 5, 6, and 9 show high F1-scores above 0.9967, with minor errors due to visual similarities, such as curved strokes in Classes 5 and 6. Class 3 has the lowest F1-score (0.9950), reflecting slight challenges in distinguishing it from Class 6. These results confirm the model’s robustness across all classes.

\subsection{Training and Validation Performance}
\label{subsec:train_val_performance}

\begin{figure}[!h]
\centering
\includegraphics[width=\columnwidth]{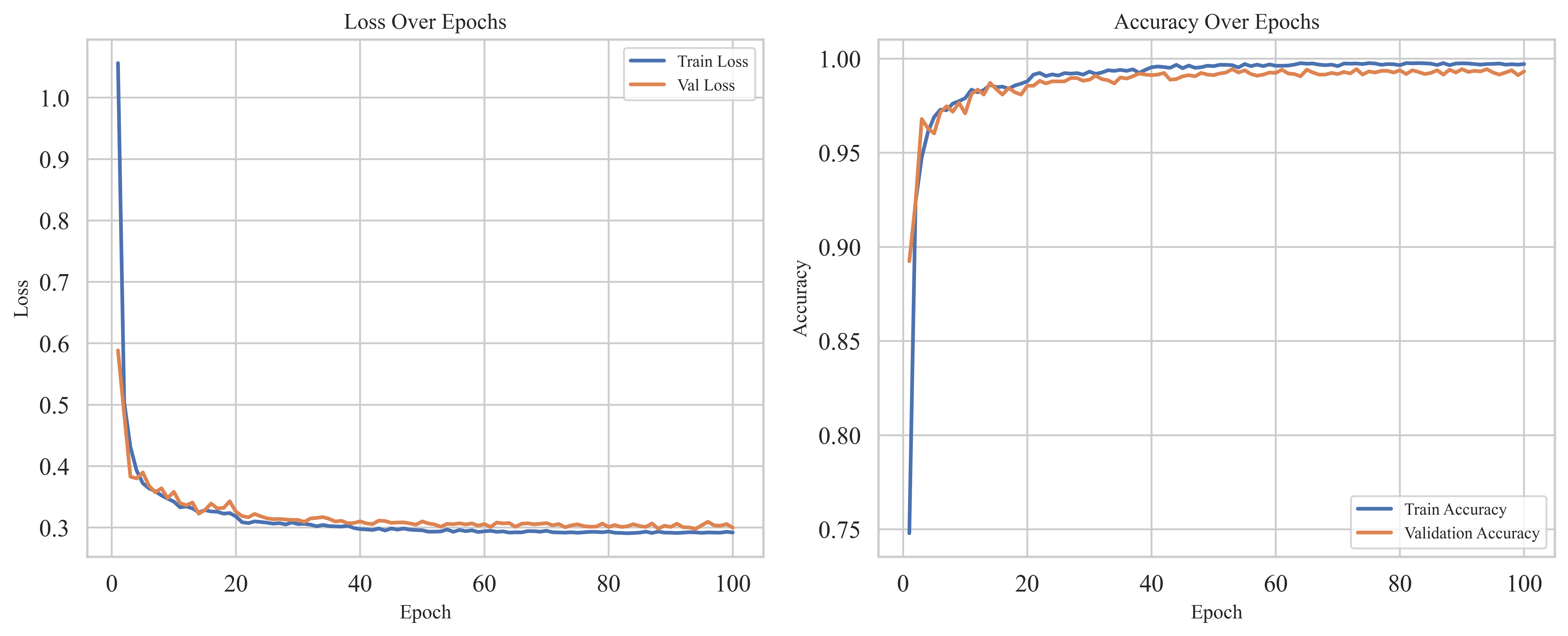}
\caption{Training and validation loss and accuracy curves over 100 epochs.}
\label{fig:train_val_plot}
\end{figure}

Figure~\ref{fig:train_val_plot} presents the training and validation loss and accuracy curves over 100 epochs for the optimal hyperparameter configuration (Section~\ref{subsec:hyperparam_results}). The training accuracy reaches 99.71\%, with validation accuracy closely tracking at 99.32\%, and validation loss at 0.2999, indicating stable convergence and minimal overfitting, underscoring its generalization capabilities, supported by the data augmentation and optimization strategies (Section~\ref{sec:experiments}).

\subsection{Hyperparameter Optimization Outcomes}
\label{subsec:hyperparam_results}

The optimal hyperparameter configuration, determined through grid search (Section~\ref{subsec:hyperparameter_tuning}), is presented in Table~\ref{tab:best_config}. The configuration includes a learning rate of $10^{-3}$, batch size of 32, dropout rate of 0.0, label smoothing of 0.05, and quantum circuit depth of 5. This setup optimizes the balance between classical and quantum components, contributing to the high test accuracy.  The absence of dropout suggests sufficient regularization through label smoothing and weight decay, while the deeper quantum circuit enhances feature discrimination.

\begin{table}[!h]
\caption{Optimal Hyperparameter Configuration}
\label{tab:best_config}
\centering
\begin{tabular}{l c}
\toprule
\textbf{Hyperparameter} & \textbf{Optimal Value} \\
\midrule
Learning Rate ($\eta$) & $10^{-3}$ \\
Batch Size ($B$) & 32 \\
Dropout Rate ($p$) & 0.0 \\
Label Smoothing ($\alpha$) & 0.05 \\
Quantum Depth ($q_d$) & 5 \\
\bottomrule
\end{tabular}
\end{table}

\subsection{Confusion Matrix Analysis}
\label{subsec:confusion_matrix}

\begin{figure}[!h]
\centering
\includegraphics[width=\columnwidth]{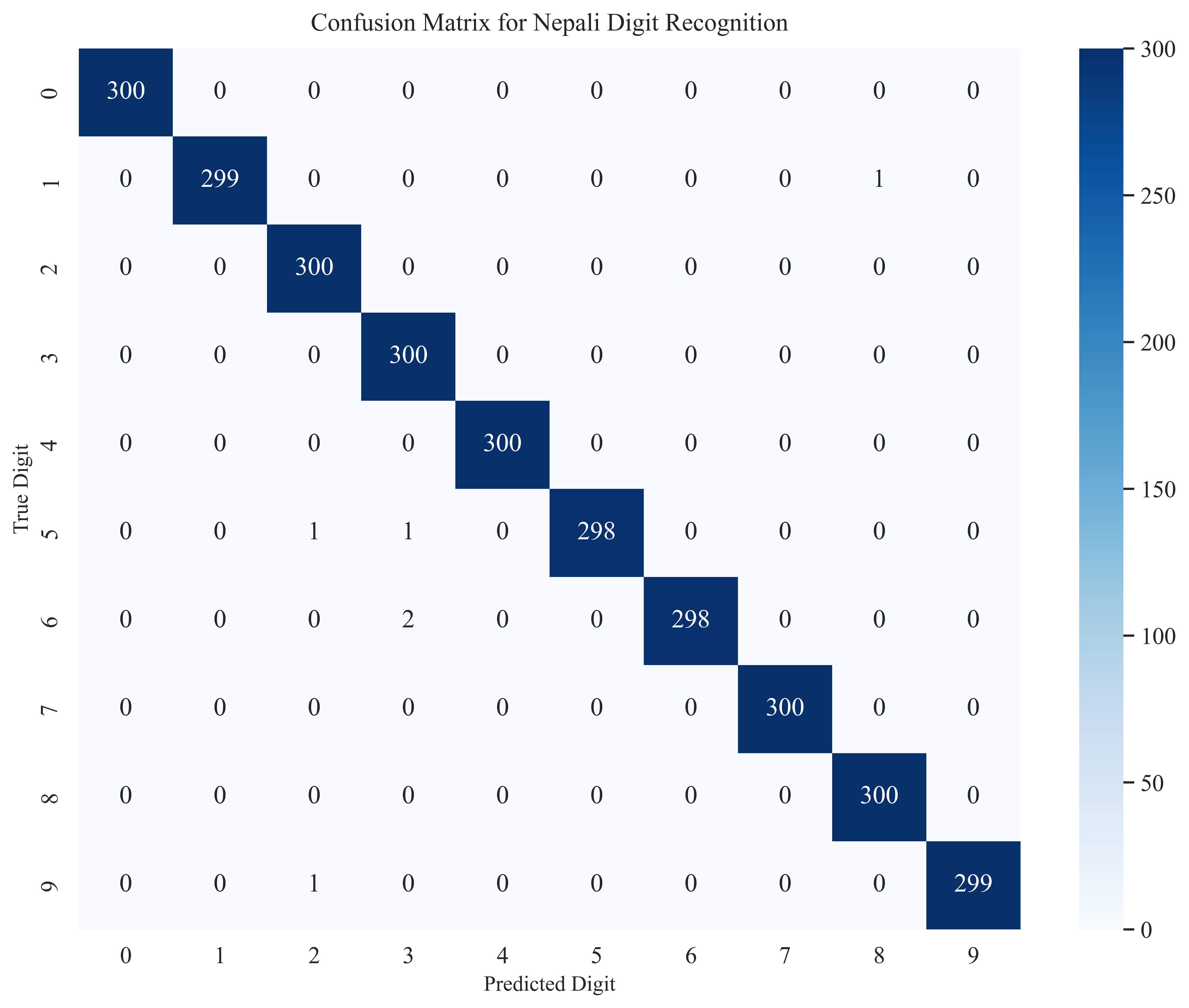}
\caption{Confusion matrix for Devanagari digit classification.}
\label{fig:conf_matrix}
\end{figure}

The confusion matrix, visualized as a heatmap in Figure~\ref{fig:conf_matrix}, maps true labels against predicted labels for the 10-digit classes, with 300 images per class. The matrix reveals only 6 misclassifications out of 3,000 test images, consistent with the 99.80\% test accuracy. Key observations include:

\begin{itemize}
    \item \textbf{Classes 0, 2, 3, 4, 7, 8}: Perfect classification with 300 correct predictions each.
    \item \textbf{Class 1}: 299 correct, with one misclassified as Class 8.
    \item \textbf{Class 5}: 298 correct, with misclassifications as Class 2 (1) and Class 3 (1).
    \item \textbf{Class 6}: 298 correct, with two misclassified as Class 3.
    \item \textbf{Class 9}: 299 correct, with one misclassified as Class 2.
\end{itemize}

The concentration of misclassifications around Classes 5 and 6 suggests visual similarities in their Devanagari forms, such as curved strokes. Despite these minor errors, the matrix confirms the model’s exceptional robustness across all classes.


\section{Discussion}
\label{sec:discussion}

This section contextualizes the results of hybrid quantum-classical models with 10 qubits on the Nepali Quantum MNIST dataset, comparing them with classical baselines, analyzing performance characteristics, and identifying limitations and future research directions.

\subsection{Comparison with Classical Methods}
\label{subsec:comparison}

Our hybrid quantum-classical models achieve test accuracies ranging from 99.47\% (4 qubits) to 99.80\% (10 qubits) on the DHCD digit subset, surpassing traditional CNNs, which typically attain 95--98\% for Devanagari digits~\cite{acharya2015}, and approaching MNIST benchmarks (e.g., 99.79\%~\cite{cirecsan2012}). An ablation study replacing the VQC with a classical linear layer yielded a test accuracy of 99.03\% and an average F1-score of 0.9912, affirming the value of the quantum component. Table~\ref{tab:ablation}Table~\ref{tab:ablation} highlights the hybrid architecture’s advantage, where quantum feature transformation enhances class separability and generalization across all qubit counts.

\begin{table}[!h]
\caption{Classical vs. Hybrid Model Performance}
\label{tab:ablation}
\centering
\begin{tabular}{l c}
\toprule
\textbf{Model} & \textbf{Test Accuracy (\%)}\\
\midrule
CNN & 99.03 \\
CNN + VQC(4 qubits) & 99.47 \\
CNN + VQC(6 qubits) & 99.67 \\
CNN + VQC(8 qubits) & 99.70 \\
CNN + VQC(10 qubits) & \textbf{99.80} \\
\bottomrule
\end{tabular}
\end{table}

\subsection{Performance Insights}
\label{subsec:insights}

The models achieve per-class F1-scores from 0.9850 to 1.0000, with misclassifications decreasing from 16 (4 qubits) to 6 (10 qubits) out of 3,000 test images. Figure~\ref{fig:conf_matrix} and Table~\ref{tab:per_class_metrics} demonstrate robust predictions across handwriting styles, with most errors occurring between visually similar digits (e.g., Classes 3 and 6). The VQC’s entanglement and quantum parallelism likely enhance discrimination of subtle variations compared to classical methods~\cite{biamonte2017}. The 10-qubit model’s near-perfect classification (99.80\% accuracy) underscores the benefit of increased qubit counts.

The hybrid architecture remains computationally efficient, with quantum circuits containing 40--100 trainable parameters and total model parameters under 94,000, significantly lighter than classical models like VGG or ResNet (20--100 million parameters). This makes our approach suitable for resource-constrained environments.

\subsection{Limitations and Future Work}
\label{subsec:limitations}

While the model achieves impressive accuracy, several limitations persist:
\begin{itemize}
    \item \textbf{Scope Restriction:} The study focuses exclusively on Devanagari digits. Extending to the full character set (including 36 consonants) is a natural next step but requires more computational and data resources.
    \item \textbf{Quantum Simulation Only:} All quantum operations were simulated using PennyLane~\cite{bergholm2018}. Real-world deployment on noisy quantum hardware may affect stability and accuracy~\cite{preskill2018}.
    \item \textbf{Limited Qubit Count:} The current architecture is constrained to 10 qubits. Exploring deeper or wider circuits with more qubits could improve expressive power, but will depend on hardware availability~\cite{cerezo2021}.
\end{itemize}

Future research will explore:
\begin{enumerate}
    \item Scaling to full Devanagari characters and other regional scripts.
    \item Deployment on quantum hardware (e.g., IBM Q, Rigetti) to assess noise robustness.
\end{enumerate}

This work establishes a robust foundation for extending QML to linguistically diverse and underrepresented writing systems.


\section{Conclusion}
\label{sec:conclusion}

We presented the first hybrid quantum-classical model for Devanagari handwritten digit recognition, achieving 99.80\% test accuracy and avearage per-class F1-scores of 0.9980 on the DHCD. By combining a CNN-based feature extractor with a compact 10-qubit VQC, the model leverages quantum superposition and entanglement to outperform classical baselines while maintaining low parameter complexity. This approach introduces a novel QML benchmark for the Devanagari script, with only 6 errors in 3,000 test samples. Beyond recognition accuracy, our model demonstrates the feasibility and promise of QML in regional, data-constrained, and multilingual contexts.


\bibliography{references} 

\begin{thebibliography}{10}
\providecommand{\url}[1]{#1}
\csname url@samestyle\endcsname
\providecommand{\newblock}{\relax}
\providecommand{\bibinfo}[2]{#2}
\providecommand{\BIBentrySTDinterwordspacing}{\spaceskip=0pt\relax}
\providecommand{\BIBentryALTinterwordstretchfactor}{4}
\providecommand{\BIBentryALTinterwordspacing}{\spaceskip=\fontdimen2\font plus
\BIBentryALTinterwordstretchfactor\fontdimen3\font minus \fontdimen4\font\relax}
\providecommand{\BIBforeignlanguage}[2]{{%
\expandafter\ifx\csname l@#1\endcsname\relax
\typeout{** WARNING: IEEEtran.bst: No hyphenation pattern has been}%
\typeout{** loaded for the language `#1'. Using the pattern for}%
\typeout{** the default language instead.}%
\else
\language=\csname l@#1\endcsname
\fi
#2}}
\providecommand{\BIBdecl}{\relax}
\BIBdecl

\bibitem{lecun2002}
Y.~LeCun, L.~Bottou, Y.~Bengio, and P.~Haffner, ``Gradient-based learning applied to document recognition,'' \emph{Proceedings of the IEEE}, vol.~86, no.~11, pp. 2278--2324, 2002.

\bibitem{acharya2015}
S.~Acharya, A.~K. Pant, and P.~K. Gyawali, ``Deep learning based large scale handwritten devanagari character recognition,'' in \emph{2015 9th International Conference on Software, Knowledge, Information Management and Applications (SKIMA)}.\hskip 1em plus 0.5em minus 0.4em\relax IEEE, 2015, pp. 1--6.

\bibitem{biamonte2017}
J.~Biamonte, P.~Wittek, N.~Pancotti, P.~Rebentrost, N.~Wiebe, and S.~Lloyd, ``Quantum machine learning,'' \emph{Nature}, vol. 549, no. 7671, pp. 195--202, 2017.

\bibitem{preskill2018}
J.~Preskill, ``Quantum computing in the nisq era and beyond,'' \emph{Quantum}, vol.~2, p.~79, 2018.

\bibitem{xia2020}
R.~Xia and S.~Kais, ``Hybrid quantum-classical neural network for calculating ground state energies of molecules,'' \emph{Entropy}, vol.~22, no.~8, p. 828, 2020.

\bibitem{krizhevsky2017}
A.~Krizhevsky, I.~Sutskever, and G.~E. Hinton, ``Imagenet classification with deep convolutional neural networks,'' \emph{Communications of the ACM}, vol.~60, no.~6, pp. 84--90, 2017.

\bibitem{aneja2019}
A.~Vidwans and S.~Aneja, ``Transfer learning using cnn for handwritten devanagari character recognition,'' in \emph{2019 1st International Conference on Advances in Information Technology (ICAIT)}.\hskip 1em plus 0.5em minus 0.4em\relax IEEE, 2019, pp. 293--296.

\bibitem{cerezo2021}
M.~Cerezo, A.~Arrasmith, R.~Babbush, S.~C. Benjamin, S.~Endo, K.~Fujii, J.~R. McClean, K.~Mitarai, X.~Yuan, L.~Cincio \emph{et~al.}, ``Variational quantum algorithms,'' \emph{Nature Reviews Physics}, vol.~3, no.~9, pp. 625--644, 2021.

\bibitem{mitarai2018}
K.~Mitarai, M.~Negoro, M.~Kitagawa, and K.~Fujii, ``Quantum circuit learning,'' \emph{Physical Review A}, vol.~98, no.~3, p. 032309, 2018.

\bibitem{phillipson2023}
F.~Phillipson, N.~Neumann, and R.~Wezeman, ``Classification of hybrid quantum-classical computing,'' in \emph{International Conference on Computational Science}.\hskip 1em plus 0.5em minus 0.4em\relax Springer, 2023, pp. 18--33.

\bibitem{abbas2021}
A.~Abbas, D.~Sutter, C.~Zoufal, A.~Lucchi, A.~Figalli, and S.~Woerner, ``The power of quantum neural networks,'' \emph{Nature Computational Science}, vol.~1, no.~6, pp. 403--409, 2021.

\bibitem{zaman2024}
K.~Zaman, T.~Ahmed, M.~A. Hanif, A.~Marchisio, and M.~Shafique, ``A comparative analysis of hybrid-quantum classical neural networks,'' in \emph{World Congress in Computer Science, Computer Engineering \& Applied Computing}.\hskip 1em plus 0.5em minus 0.4em\relax Springer, 2024, pp. 102--115.

\bibitem{bergholm2018}
V.~Bergholm, J.~Izaac, M.~Schuld, C.~Gogolin, S.~Ahmed, V.~Ajith, M.~S. Alam, G.~Alonso-Linaje, B.~AkashNarayanan, A.~Asadi \emph{et~al.}, ``Pennylane: Automatic differentiation of hybrid quantum-classical computations,'' \emph{arXiv preprint arXiv:1811.04968}, 2018.

\bibitem{glorot2010}
X.~Glorot and Y.~Bengio, ``Understanding the difficulty of training deep feedforward neural networks,'' in \emph{Proceedings of the thirteenth international conference on artificial intelligence and statistics}.\hskip 1em plus 0.5em minus 0.4em\relax JMLR Workshop and Conference Proceedings, 2010, pp. 249--256.

\bibitem{szegedy2016}
C.~Szegedy, V.~Vanhoucke, S.~Ioffe, J.~Shlens, and Z.~Wojna, ``Rethinking the inception architecture for computer vision,'' in \emph{Proceedings of the IEEE conference on computer vision and pattern recognition}, 2016, pp. 2818--2826.

\bibitem{loshchilov2017}
I.~Loshchilov and F.~Hutter, ``Decoupled weight decay regularization,'' \emph{arXiv preprint arXiv:1711.05101}, 2017.

\bibitem{cirecsan2012}
D.~Ciregan, U.~Meier, and J.~Schmidhuber, ``Multi-column deep neural networks for image classification,'' in \emph{2012 IEEE conference on computer vision and pattern recognition}.\hskip 1em plus 0.5em minus 0.4em\relax IEEE, 2012, pp. 3642--3649.

\end{thebibliography}

\end{document}